\documentclass[12pt]{article}

\begin{document}

\title{Petrov types of slowly rotating fluid balls}
\author{Gyula Fodor$^{1,2}$ and Zolt\'{a}n Perj\'{e}s$^1$ \\
$1$ \ \ KFKI Research Institute for Particle and Nuclear Physics, \\
H-1525, Budapest 114, P.O.B. 49, Hungary \\
$2$ \ \ Department of Physics, Waseda University, 3-4-1 Okubo, \\
Shinjuku, Tokyo 169-8555, Japan}
\maketitle

\begin{abstract}
Circularly rotating axisymmetric perfect fluid space-times are investigated
to second order in the small angular velocity. The conditions of various
special Petrov types are solved in a comoving tetrad formalism. A
number of theorems are stated on the possible Petrov types of various fluid
models. It is shown that Petrov type II solutions must reduce to the de
Sitter spacetime in the static limit. Two space-times with a physically
satisfactory energy-momentum tensor are investigated in detail. For the
rotating incompressible fluid, it is proven that the Petrov type cannot be
D. The equation of the rotation function $\omega $ can be solved for the
Tolman type IV fluid in terms of quadratures. It is also shown that the
rotating version of the Tolman IV space-time cannot be Petrov type D.
\end{abstract}

%
%
%
%
%

\section{Introduction}

Following Schwarzschild's discovery of the static incompressible interior
solution in 1916, relentless efforts have been made to find a rotating
generalization. Improved equations of state for the perfect fluid were
imported in general relativity, leaving, however little clue for how to
achieve this goal. Less than that, the supply of even the physically
acceptable \emph{nonrotating} solutions is scarce \cite{Delgaty}. The common
approach to solve the field equations of the problem is to make some \textit{%
a priory} assumption on the properties of the desired spacetime, \textit{i.\
e.},\ make an \textit{ansatz}, and hope that solutions with the
corresponding property do exist. More often than not, after considerable
effort one gets a highly unphysical solution in this procedure, if any
solution turns out to exist at all. A natural way to decide the
acceptability of the various ans\"{a}tze is to check whether or not they
remain valid in the slowly rotating limit. Since static spherically
symmetric spacetimes are either Petrov type D or 0, a widely studied ansatz
is to assume that the rotating fluid also belongs to some special Petrov
class. The aim of this paper is to investigate whether or not these classes
contain physically acceptable perfect fluid spacetimes. It is natural to
assume, that such configurations have a well behaving slow rotation limit.

We investigate axisymmetric stationary perfect fluid spacetimes in circular
rotation, \textsl{i.e.,} with the fluid velocity vector lying in the plane
of the timelike and angular Killing vectors, $\partial /\partial t$ and $%
\partial /\partial \varphi $ respectively. The metric of the spacetime is
written in the form 
\begin{equation}
ds^2=\tilde{X}^2dt^2-\tilde{Y}^2dr^2-\tilde{Z}^2\left[ d\vartheta ^2+\sin
^2\vartheta (d\varphi -\omega dt)^2\right] \ ,  \label{dsgen}
\end{equation}
where $\tilde{X}$, $\tilde{Y}$, $\tilde{Z}$ and $\omega $ are functions of
the coordinates $r$ and $\vartheta .$ We choose the time-translation Killing
vector $\partial /\partial t$ such that it becomes asymptotically
nonrotating at spacelike infinity. With this choice, the \textit{angular
velocity} $\Omega $ of the fluid is defined by the components of the fluid
velocity vector $u^\mu $ as follows, 
\begin{equation}
u^\varphi =\Omega u^t\ .
\end{equation}
We further assume that the fluid is in \textit{rigid rotation}, i.\ e. $%
\Omega $ is a constant. This $\Omega $ is the parameter which is small in
the slow rotation limit, and following Hartle's work \cite{Hartle}, we will
expand quantities in powers of $\Omega $. For rigidly rotating space-times,
a \textit{comoving coordinate system}, where $u^\varphi =0,$ can be arranged
by a linear transformation of the angular coordinate $\varphi
\longrightarrow \varphi -\Omega t$. Since in this work we are interested
only in the interior fluid region, we will perform our calculations in the
comoving system.

We focus on the Petrov types that a slowly rotating fluid ball might have.
We use a tetrad, constructed specifically for this task, representing the
space-time to the desired order in the angular velocity $\Omega $. In this
framework, definite statements can be made about the Petrov type of the
field of a given order in $\Omega$, or about the impossibility of certain
Petrov types \textit{to any order} in $\Omega .$ Nevertheless, the slow
rotation approximation can be applied to give information about numerous
physical properties, even about the possible existence of an asymptotically
flat exterior vacuum region \cite{bfmp}.

Since the system behaves the same way under a reversal in the direction of
rotation as under a reversal in the direction of time, when expanding the
components of the metric (\ref{dsgen}) in powers of the angular velocity
parameter $\Omega $ one finds that $\omega $ contains only odd powers, while 
$\tilde{X}$, $\tilde{Y}$ and $\tilde{Z}$ contain solely even powers of $%
\Omega $. In this paper we are interested in effects of up to $\Omega ^2$
order, and hence we consider the function $\omega $ to be proportional to
the angular velocity parameter $\Omega $, while we allow $\Omega ^2$ terms
in the diagonal components of the metric (\ref{dsgen}). The metric of a
slowly rotating fluid ball can be written in the form \cite{Hartle} 
\begin{eqnarray}
ds^2 &=&(1+2h)X^2dt^2-(1+2m)Y^2dr^2  \nonumber \\
&&-(1+2k)Z^2\left[ d\vartheta ^2+\sin ^2\vartheta \left( d\varphi -\omega
dt\right) ^2\right] \ .  \label{ds}
\end{eqnarray}
Here the functions $X$, $Y$ and $Z$ depend only on the radial coordinate $r$%
, determining the spherically symmetric basis solution, while $\omega $, $h$%
, $m$ and $k$ are functions of both $r$ and $\vartheta $. The potential $%
\omega $ is small to first order in the angular velocity $\Omega $, but $h$, 
$m$ and $k$ are second order small quantities. There are two minor
differences between our metric form (\ref{ds}) and the corresponding formula
in the paper of Hartle \cite{Hartle}. The first is that we do not use the
radial gauge $Z=r$, since that choice is technically inconvenient for
certain spherically symmetric perfect fluid exact solutions. The second
difference is that our definition of the second order small quantity $m$
differs by a zeroth order factor. This choice is only to make our equations
shorter, especially because the local mass function in the denominator of
Hartle's definition takes a more complicated form in a general radial gauge $%
Z\not=r$.

Let us consider a freely falling observer, with velocity vector $v^\mu $,
who has zero impact parameter and consequently zero angular momentum. In the
coordinate system where $\partial /\partial t$ corresponds to the timelike
Killing vector which is nonrotating at spacelike infinity, $v_\varphi =0$,
and the function $\omega $ agrees with the angular velocity $v^\varphi
/v^t=g^{\varphi t}/g^{tt}=-g_{\varphi t}/g_{\varphi \varphi }$ of this
observer. Hence $\omega $ can be interpreted as the rate of rotation of the
local inertial frame with respect to the distant stars\cite{Hartle}, or in
other words, the cumulative dragging of initial frames\cite{Miller}. In
comoving coordinates, the rotation potential $\omega $ represents the
coordinate angular velocity of the fluid element at $\left( r,\vartheta
\right) $ measured by a freely falling observer to first order in $\Omega $.

We next briefly recapitulate Hartle's arguments leading to an unambiguous
choice of the coordinate system. For the full details, \textit{\ cf.} \cite
{Hartle}. We start from a known spherically symmetric perfect fluid
solution, which is described by the functions $X$, $Y$ and $Z$. Calculating
the field equations to \textit{first order} in the small angular velocity
parameter $\Omega $, we get only one independent relation \cite{brco} \cite
{cobr}. The $(t,\varphi )$ component of Einstein equation gives a second
order partial differential equation for $\omega $. In the comoving
coordinate system this equation takes the form 
\begin{equation}
\frac X{YZ^2}\frac \partial {\partial r}\left( \frac{Z^4}{XY}\frac{\partial
\omega }{\partial r}\right) +4\frac{XZ}Y\left[ \frac d{dr}\left( \frac 1{XY}%
\frac{dZ}{dr}\right) \right] \omega +\frac 1{\sin ^3\vartheta }\frac
\partial {\partial \vartheta }\left( \sin ^3\vartheta \frac{\partial \omega 
}{\partial \vartheta }\right) =0\ .  \label{E03}
\end{equation}
Expansion of (\ref{E03}) in vector spherical harmonics yields the angular
behavior of the solution in the form 
\begin{equation}
\omega =\sum_{l=1}^\infty \omega _l\left[ -\frac 1{\sin \vartheta }\frac{%
dP_l(\cos \vartheta )}{d\vartheta }\right] \ ,
\end{equation}
where the functions $\omega _l$ depend only on the radial coordinate $r$,
and $P_l$ is the Legendre polynomial of order $l$. The equations for the
coefficients $\omega _l$ with different values of $l$ decouple. Taking into
account the matching conditions at the surface of the fluid ball, one can
show that for $l>1$ the functions $\omega _l$ cannot be regular both at the
center of the fluid and at infinity. It follows from the asymptotic flatness
of the exterior spacetime region that $\omega $ cannot depend on the angular
coordinate $\vartheta $. Consequently, the rotation potential $\omega $ is a
function of the radial coordinate $r$ alone, even in the fluid region,
satisfying a second-order ordinary linear differential equation.

Since the expansion of $\omega $ in the angular velocity parameter cannot
contain $\Omega ^2$ terms, the solution of (\ref{E03}) will remain valid to
second order as well. After solving the first order condition for the
function $\omega $, proceeding to second order in the angular velocity
parameter $\Omega $, the components of the Einstein equation give a coupled
linear inhomogeneous system of partial differential equations for the
functions $h$, $m$ and $k$. The inhomogeneous terms in the equations are
proportional to $\omega ^2$ and its derivatives. The solution of this system
can be written in the form of an expansion in spherical harmonics, 
\begin{equation}
h=\sum_{l=0}^\infty h_lP_l(\cos \vartheta )\ ,
\end{equation}
and similarly for $m$ and $k$. However, quantities with different $l$
decouple, and the equations for $l>2$, being homogeneous, do not include $%
\omega $. As a result, all $h_l$, $m_l$ and $k_l$ must vanish for $l>2$,
since otherwise they would correspond to a static but not spherically
symmetric configuration in the $\omega =0$ case. Thus the second order small
functions can be written in the form 
\begin{eqnarray}
h &=&h_0+h_2P_2(\cos \vartheta )\ ,  \nonumber \\
m &=&m_0+m_2P_2(\cos \vartheta )\ , \\
k &=&k_2P_2(\cos \vartheta )\ ,  \nonumber
\end{eqnarray}
where $h_0$, $h_2$, $m_0$, $m_2$ and $k_2$ are functions of $r$. The freedom
in the choice of radial coordinate was used to set the $\vartheta $
independent part of $k$ to zero.

In Ref.\ \cite{Hartle}, Hartle writes out the detailed form of the equations
describing the second order rotational perturbations in the $Z=r$ gauge, and
gives a procedure to determine the binding energy, the baryon number change,
and the ellipticity of the fluid surface. However, this procedure involves
numerical integration of a system of ordinary differential equations.
Instead, in our paper, we focus on what can be said about the general
physical properties of slowly rotating fluid bodies by analytical methods,
without trying to solve the perturbation equations. There is an extensive
literature on numerical simulations of slowly rotating bodies for various
types of fluids. These include neutron stars and supermassive stars \cite
{HarTho}, incompressible fluids \cite{ChaMil}, polytropes \cite{Hartle} and
realistic neutron matter equations of state \cite{WGW}.

In Sec.\ 2., we establish a tetrad formalism for the Petrov classification.
We also prove a theorem showing that physically realistic slowly rotating
perfect fluid spacetimes cannot be Petrov type II. Case studies for various
equations of state are presented in the rest of the paper. In Sec.\ 3., we
establish a theorem on incompressible fluids. This theorem indicates that
circularly rotating states should be found in the algebraically general
class. For Tolman IV fluids, in Sec.\ 4., we find the rotation function $%
\omega $ in terms of quadratures, and we show that these fluids cannot be
Petrov type D.

The rotation function $\omega $ can be written down in terms of elementary
functions for the rotating Whittaker space-time. Although the equation of
state of the Whittaker fluid implies that the density decreases towards the
center of the fluid ball, this class merits special attention. Among the
rotating states of the Whittaker fluid, there is the exactly known Petrov
type D Wahlquist solution. The rotating Whittaker fluid is dealt with in 
\cite{bfmp} and \cite{bfp}.

\section{The Petrov types}

To quadratic order in the angular velocity parameter $\Omega $, the
nonvanishing components of a comoving tetrad can be chosen for the metric (%
\ref{ds}) as 
\begin{eqnarray}
&&e_0^t=\left( 1+\frac 12\omega ^2\frac{Z^2}{X^2}\sin ^2\vartheta -h\right)
\frac 1X\ ,  \nonumber \\
&&e_1^r=\frac{1-m}Y\ ,\ \ e_2^\vartheta =\frac{1-k}Z\ ,\ \ e_3^t=\omega
\frac Z{X^2}\sin \vartheta \ ,  \label{tetr} \\
&&e_3^\varphi =\left( -1+\frac 12\omega ^2\frac{Z^2}{X^2}\sin ^2\vartheta
+k\right) \frac 1{Z\sin \vartheta }\ .  \nonumber
\end{eqnarray}
The correctness of these expressions to the required order can be shown by
checking that $(e_a^\mu e_b^\nu g_{\mu \nu })=\mathrm{diag}(1,-1,-1,-1)$ up
to $\Omega ^2$ terms, where Roman and Greek labels denote tetrad and
spacetime indices, respectively, and $g_{\mu \nu }$ are the components of
the metric (\ref{ds}).

Using the tetrad components (\ref{tetr}), we will compute the Ricci rotation
coefficients, the tetrad components of the Riemann, Weyl and Einstein
tensors, and also some expressions, which are vanishing for certain Petrov
types. All these quantities are polynomial expressions in the metric
components $g_{\mu \nu }$, the tetrad vector components $e_a^\mu $, and
their coordinate derivatives. Since the tetrad vector components will be
multiplied only by terms which have a regular $\Omega =0$ limit, we get the
correct results to second order in $\Omega $ even if we use expressions for
the tetrad, which are only correct to the same order.\footnote{%
An alternative approach would be to consider temporarily (\ref{ds}) to be
valid to arbitrary order in the angular velocity $\Omega $, calculate the
exact tetrad vector components, the Riemann tensor and other necessary
quantities, and finally expand the result to second order in $\Omega $. This
procedure gives exactly the same results as the simpler one, but because of
the appearance of complicated denominator terms it would be a tedious work
even for modern computer algebra systems. An even more cautious way of
calculation, leading to the same results once again, would be to compute the
coordinate components of the Riemann tensor, and take the tetrad components
only in the end. However, this approach would have a similar problem with
the components of the inverse metric $g^{\mu \nu }$, namely, whether or not
one calculates them only to second order from the beginning.}

The electric and magnetic curvature components are defined in terms of the
tetrad components of the Weyl tensor as follows \cite{fmp}: 
\[
\begin{array}{lll}
E_1=C_{1010}\ , & E_2=C_{2020}\ , & E_3=C_{1020}\ , \\ 
H_1=^{\;*}\!\!C_{1010}\ , & H_2=^{\;*}\!\!C_{2020}\ , & H_3=^{\;*}\!%
\!C_{1020}\ .
\end{array}
\]
Because of the symmetries of the configuration, these six components are the
only independent components of the Weyl tensor. A Newman-Penrose
frame is readily defined by $\mathbf{l}=(e_0+e_3)/\sqrt{2},\ \mathbf{n}%
=(e_0-e_3)/\sqrt{2},\ \mathbf{m}=(e_1+ie_2)/\sqrt{2}$. The nonvanishing Weyl
spinor components are 
\begin{eqnarray}
\Psi _0 &=&\frac 12[E_2-E_1-2H_3+i(H_1-H_2-2E_3)]\ ,  \label{Psi0} \\
\Psi _2 &=&\frac 12[E_1+E_2-i(H_1+H_2)]\ ,  \label{Psi2} \\
\Psi _4 &=&\frac 12[E_2-E_1+2H_3+i(H_1-H_2+2E_3)]\ .
\end{eqnarray}

The Petrov type of the spacetime can be determined by studying the
properties of the Weyl spinor components. If $\Psi _0=\Psi _2=\Psi
_4=0$ the spacetime is conformally flat, consequently it is the nonrotating
interior Schwarzschild solution\cite{Coll}. If $\Psi _0=\Psi _4=0$ but $\Psi
_2\not=0$, one can apply a tetrad rotation around the vector $\mathbf{l}$,
which makes $\mathbf{m}$ into $\mathbf{m}+a\mathbf{l}$. Then the new nonzero
Weyl spinor components become $\tilde{\Psi}_4=6(a^{*})^2\Psi _2$, $\tilde{%
\Psi}_3=3a^{*}\Psi _2$ and $\tilde{\Psi}_2=\Psi _2$, and the Petrov type is
determined by the number of the distinct roots of the equation $0=\tilde{\Psi%
}_4b^4+4\tilde{\Psi}_3b^3+6\tilde{\Psi}_2b^2\equiv 6\Psi _2b^2(a^{*}b+1)^2$.
This equation has two double roots, which shows that the spacetime is of
type D. If $\Psi _4\not=0$ the Petrov type is determined by the
number of the distinct roots of the algebraic equation 
\begin{equation}
\Psi _4b^4+6\Psi _2b^2+\Psi _0=0  \label{peteq}
\end{equation}
for the parameter $b$. If $\Psi _4=0$ but $\Psi _0\not=0$ one can apply a
reversal in the $e_3$ direction to interchange the frame vectors $\mathbf{l}$
and $\mathbf{n}$, and consequently interchange $\Psi _4$ and $\Psi _0$ as
well. Type III solutions are excluded because $\Psi _3$ and $\Psi _1$
vanish. The type is N if $\Psi _0=\Psi _2=0$ but $\Psi _4\not=0$. It has
been shown in \cite{fmp} that all axistationary type N perfect fluid
solutions can be interpreted as vacuum solutions with a negative
cosmological constant. In the case $\Psi _4\not=0$ the Petrov type is II
if and only if $\Psi _0=0$ but $\Psi _2\not=0$. Assuming $\Psi _4\not=0$, 
the Petrov type is D if and only if $9\Psi _2^2=\Psi_0\Psi _4\not=0$. 
For our generic comoving tetrad, the real and imaginary
parts of the equation $9\Psi _2^2=\Psi _0\Psi _4$ yield 
\begin{eqnarray}
2E_1^2+5E_1E_2+2E_2^2-E_3^2-2H_1^2-5H_1H_2-2H_2^2+H_3^2 &=&0\ ,
\label{PetD2} \\
4E_1H_1+5E_1H_2+5E_2H_1+4E_2H_2-2E_3H_3 &=&0\ .  \label{PetD1}
\end{eqnarray}
These conditions characterize slowly rotating fields. For finite angular
velocities, the conditions for higher-order terms may not be met any more.
By the generic properties of the power series expansion of differentiable
functions, however, any statement about the \textit{nonexistence} of a
certain type slowly rotating fluid will hold in the exact sense.

In the spherically symmetric limit, when $\Omega =0$, it is easy to check
that $\Psi _0=\Psi _4=-3\Psi _2$, independently of the equation of state.
There are only two possible Petrov classes then. If all $\Psi _i$ are zero,
the metric is the conformally flat Petrov type 0 incompressible interior
Schwarzschild solution. All other spherically symmetric perfect fluid
spacetimes are Petrov type D, since $9\Psi _2^2=\Psi _0\Psi _4\not=0$
holds for them. This has important consequences for rotating Petrov N and
Petrov II fluids, and also for the $\Psi _0=\Psi _4=0$ Petrov D subclass.
Since $\Psi _4$ or $\Psi _0$ vanishes in all of these cases, in the $\Omega
=0$ limit all the Weyl spinor components must go to zero, and consequently
these spacetimes must reduce to the incompressible interior Schwarzschild
solution in the nonrotating limit. This is already a severe limitation, but
as we will see shortly, higher order conditions on the slowly rotating fluid
state will pose further restrictions, which make these classes irrelevant
for the study of rotating isolated bodies.

\textbf{Lemma 1:} Circularly and rigidly rotating perfect fluids with $\Psi
_0=0$ must reduce to the de Sitter space-time in the slow-rotation limit.

\textit{Proof: } Using (\ref{Psi0}), the real and imaginary parts of the
condition $\Psi _0=0$ give 
\begin{eqnarray}
E_2-E_1-2H_3 &=&0\ ,  \label{Eeq} \\
H_1-H_2-2E_3 &=&0\ .  \label{Heq}
\end{eqnarray}
We can prove the lemma without assuming the existence of an
asymptotically flat region, i.\ e.\ when $\omega $ may also depend on $%
\vartheta $. We denote the tetrad components of the Einstein tensor by $%
G_{ab}$. As we have discussed in the previous paragraph, the spacetime must
reduce to the incompressible interior Schwarzschild solution in the
nonrotating limit, and hence the metric is of the form (\ref{ds}) with
\begin{eqnarray}
X &=&A-\cos r\ ,  \nonumber \\
Y &=&R\ ,  \label{intsch} \\
Z &=&R\sin r\ ,  \nonumber
\end{eqnarray}
where $A$ and $R$ are constants, satisfying $R>0$ and $1<A<3$. It is also
possible to get this result directly by solving the zeroth order parts of
the equations (\ref{Eeq}) and $G_{11}=G_{22}$. A linear
combination of the first order parts of $G_{03}=0$ and (\ref{Eeq}) yields a
differential equation for $\omega $ without $\vartheta $ derivatives. The
general solution of that is 
\begin{equation}
\omega =\left( \frac{2A}{\cos (r)+1}-1\right) f_1(\vartheta )+\left( \frac{2A%
}{\cos (r)-1}-1\right) f_2(\vartheta )\ ,
\end{equation}
where $f_1(\vartheta )$ and $f_2(\vartheta )$ are some functions of the $%
\vartheta $ coordinate. This $\omega $ is regular at the center of the fluid 
$r=0$ only if $f_2(\vartheta )=0$ or if $A=0$. Supposing that $A$ is nonzero,
substituting into (\ref{Heq}) the leading $\cos ^2r$ terms yield 
\begin{equation}
A\left( \sin \vartheta \frac{df_1(\vartheta )}{d\vartheta }+2\cos \vartheta
f_1(\vartheta )\right) =0\ .
\end{equation}
The solution of the second factor is proportional to $1/\sin ^2\vartheta $,
which is divergent at the rotation axis $\vartheta =0$. Consequently we must
have $A=0$. Substituting back again to (\ref{Eeq}) and (\ref{Heq}) we get
that the derivative of $f_1(\vartheta )$ must be zero, and consequently $%
\omega $ is a constant. Since $A=0$, this is the de Sitter space-time in a
rotating coordinate system. The pressure $p$ and density $\mu $ are
constants, $p=-\mu =3/R$. Thus the equation of state violates the weak
energy condition.

The Petrov type is II in the following two cases: if and only if $
\Psi _0=0$ but $\Psi _2$ and $\Psi _4$ are nonzero, or when $\Psi _4=0$ but 
$\Psi _2$ and $\Psi _0$ are nonzero. In the $\Psi _4=0$ case we
change the components of $e_3$ to $-1$ times those in (\ref{tetr}) in order
to exchange $\Psi _0$ with $\Psi _4$. From our Lemma, it follows that

\textbf{Theorem 1:} Circularly and rigidly rotating perfect fluids of Petrov
type II must reduce to the de Sitter space-time in the slow-rotation limit.

Our results also show that a slowly rotating perfect fluid with an
equation of state satisfying the weak energy condition cannot be of Petrov
type II.

Lemma 1 also applies to the $\Psi _4=\Psi _0=0$ but $\Psi _2\not=0$
Petrov type D subcase. This shows that all physically acceptable rotating
Petrov type D solutions must be in the $9\Psi _2^2=\Psi _0\Psi _4\not=0$
class. Since this case seems to be too complicated for general
investigation, we proceed with studying perfect fluids with specific
equations of states.

\section{Incompressible fluid}

The Schwarzschild metric of a non-rotating incompressible perfect fluid ball
is described by (\ref{intsch}). The pressure and density are 
\begin{equation}
\mu =\frac 3{R^2}\ ,\qquad p=\frac{3\cos r-A}{R^2(A-\cos r)}\ .
\end{equation}

We assume the existence of an exterior asymptotically vacuum region, in
which case $\omega$ cannot depend on $\theta$. Calculating to second order
in the small angular velocity parameter $\Omega$, the $(t,\varphi )$
component (\ref{E03}) of the Einstein equations gives the condition 
\begin{equation}
(A-\cos r)\sin r\frac{d ^{2}\omega }{d r^{2}}-\left( 3\cos ^{2}r-4A\cos
r+1\right) \frac{d \omega }{d r}-4A\omega \sin r=0\ .  \label{seco}
\end{equation}
There are also three other second-order field equations involving the
functions $h$, $m$ and $k$.

Using the tetrad (\ref{tetr}), we get that the vorticity is linear in the
angular velocity parameter $\Omega $, 
\begin{eqnarray}
\omega _1 &=&\frac{\cos \vartheta }{\cos r-A}\omega \ , \\
\omega _2 &=&\frac{\sin \vartheta }{2\left( A-\cos r\right) ^2}\left[
2\omega (A\cos r-1)+\sin r\left( A-\cos r\right) \frac{d\omega }{dr}\right]
\ .
\end{eqnarray}
The changes of all other rotational coefficients are small to second order.

The electric part of the Weyl tensor is quadratic in the angular velocity
parameter $\Omega $, and the magnetic part has only linear terms, 
\begin{eqnarray}
H_1 &=&\frac{\cos \vartheta }{R\left( \cos r-A\right) }\frac{d\omega }{dr}\
,\ \ H_2=-\frac 12H_1\ , \\
H_3 &=&\frac{\sin \vartheta }{2R(A-\cos r)^2}\left[ \cos r(\cos r-A)\frac{%
d\omega }{dr}+2A\omega \sin r\right]  \nonumber \\
&=&-\frac{\omega _2\cos r}{R\sin r}+\frac{\omega \sin \vartheta }{R\sin
r(A-\cos r)}\ .
\end{eqnarray}
In getting $H_3$, we used (\ref{seco}) to eliminate the second derivative of 
$\omega $.

Since the other subcase is excluded by Lemma 1, the Petrov type can be D if
and only if Eqs.\ (\ref{PetD2}) and (\ref{PetD1}) hold. For the slowly
rotating incompressible fluid the only condition they give is $%
H_3^2-2H_1^2-2H_2^2-5H_1H_2=0$, which implies 
\begin{equation}
\cos r(\cos r-A)\frac{d\omega }{dr}+2\omega A\sin r\ =0\ .
\end{equation}
The general solution of this equation is 
\begin{equation}
\omega =C\left( \frac A{\cos r}-1\right) ^2\ ,
\end{equation}
where $C$ is a constant. However, this $\omega $ is not a solution of (\ref
{seco}), and hence we have proven:

\textbf{Theorem 2:} A slowly and circularly rotating incompressible perfect
fluid spacetime with an asymptotically flat vacuum exterior cannot be Petrov
type~D.

\section{Tolman fluid}

A spherically symmetric perfect fluid space-time has been given in \cite
{Tolman}. The metric is described by 
\begin{eqnarray}
X^{2} &=&\frac{B^{2}\left( r^{2}+A^{2}\right) }{A^{2}} \ ,  \nonumber \\
Y^{2} &=&\frac{A^{2}+2r^{2}}{\left( R^{2}-r^{2}\right) \left(
A^{2}+r^{2}\right) }R^{2} \ , \\
Z &=&r \ ,  \nonumber
\end{eqnarray}
where $A$, $B$ and $R$ are constants. The equation of state is quadratic in
the pressure, 
\begin{equation}
R^2(A^2+2R^2)\mu=4R^4A^2p^2+R^2(2R^2+13A^2)p+6(R^2+2A^2) \ .
\end{equation}
The metric has a pleasingly simple form and describes an isolated fluid body
with a regular center, outwards decreasing density and pressure and
subluminal sound speed for appropriately chosen parameters \cite{Delgaty}.

We assume the existence of an asymptotically flat exterior vacuum region, in
which case the function $\omega$ does not depend on $\vartheta$, and the $%
(t,\varphi)$ component (\ref{E03}) of the Einstein equations gives 
\begin{eqnarray}
&&\left( r^2-R^2\right) \left( 2r^2+A^2\right) \frac{d\omega ^2}{dr^2}%
+\left( 5A^2r-4A^2\frac{R^2}r+8r^3-6rR^2\right) \frac{d\omega }{dr} 
\nonumber \\
&&\ \ \ \ \ \ +4\left( 2R^2+A^2\right) \omega =0\ .  \label{tolmom}
\end{eqnarray}
The general solution of this is 
\begin{eqnarray}
\omega &=&\frac{\sqrt[4]{2r^2+A^2}}{r^2\sqrt[4]{R^2-r^2}}\Biggl[ C_1\exp
\left( \frac 12\int \frac{P_1+2\sqrt{P_2}}{P_3}dr\right)  \nonumber \\
&&\qquad \qquad \quad +C_2\exp \left( \frac 12\int \frac{P_1-2\sqrt{P_2}}{P_3%
}dr\right) \Biggr] \ ,
\end{eqnarray}
where $C_1$ and $C_2$ are constants and 
\begin{eqnarray}
&&P_1=8r^{10}+\left( 10R^2+13A^2\right) r^8+\left( 6A^4+4A^2R^2-8R^4\right)
r^6  \nonumber \\
&&\ \ +A^2\left( A^4+4A^2R^2-4R^4\right) r^4+R^2A^4\left( A^2-6R^2\right)
r^2-2A^6R^4 \\
&&P_2=R^2r^6\left( 8R^2+A^2\right) \left( 5R^2+4A^2\right) \left(
R^2-r^2\right) \left( 2r^2+A^2\right) ^3 \\
&&P_3=r\left( r^2-R^2\right) \left( 2r^2+A^2\right)  \nonumber \\
&&\qquad \left( r^6+4r^4R^2+2A^2r^4+r^2A^4+2r^2A^2R^2-A^4R^2\right) \ .
\end{eqnarray}

Since the $\Psi _0=\Psi _4=0$ subcase is excluded by Lemma 1, the rotating
fluid can be Petrov type D only if equations (\ref{PetD2}) and (\ref{PetD1})
are satisfied. Using the tetrad (\ref{tetr}) we get that (\ref{PetD1}) holds
identically to second order in the angular velocity parameter $\Omega $.
After substituting for the second derivative of $\omega $ from (\ref{tolmom}%
) the Petrov D condition (\ref{PetD2}) yields 
\begin{eqnarray}
&&(r^2-R^2)^2(A^2+2r^2)\left( \frac{d\omega }{dr}\right)
^2+4r(r^2-R^2)(A^2+2R^2)\omega \frac{d\omega }{dr}  \label{ford1} \\
&&\ \ \ \ +4r^2\frac{(A^2+2R^2)^2}{A^2+2r^2}\omega ^2+6B^2R^2\frac{%
(A^2+2R^2)(A^2+2r^2)}{A^2(r^2+A^2)}(m_2-h_2)=0\ .  \nonumber
\end{eqnarray}
Unlike in the incompressible case, this condition now involves the second
order small quantities $m_2$ and $h_2$. The reason for the easier treatment
of the interior Schwarzschild solution was that because of its
conformal-flat nature, both the electric and magnetic parts of the Weyl
tensor were vanishing in the nonrotating limit. To be able to decide about
the Petrov type, in general, one has to consider those parts of the Einstein
equations, which are second order small in the angular velocity parameter $%
\Omega $. There are three such equations for the Tolman IV fluid. The
pressure isotropy condition $G_{22}=G_{33}$ takes the form 
\begin{eqnarray}
&&A^2r^4(r^2-R^2)(A^2+2r^2)\left( \frac{d\omega }{dr}\right)
^2-4A^2r^4(A^2+2R^2)\omega ^2  \nonumber \\
&&\quad +6B^2R^2(A^2+2r^2)^2(m_2+h_2)=0\ .  \label{ford2}
\end{eqnarray}
The condition $G_{12}=0$ gives 
\begin{equation}
r(A^2+r^2)\frac d{dr}(h_2+k_2)-A^2h_2-(A^2+2r^2)m_2=0\ ,  \label{w1}
\end{equation}
while the $P_2(\cos \vartheta )$ part of the another pressure-isotropy
condition $G_{11}=G_{22}$ yields a more lengthy second order differential
equation 
\begin{eqnarray}
&&3B^2r(r^2\!-R^2)(A^2+2r^2)\left[ (A^2+r^2)r\frac{d^2(k_2+h_2)}{dr^2}%
-(A^2+2r^2)\frac{dm_2}{dr}\right]   \nonumber \\
&&+3B^2r(4r^6+A^2r^4-2R^2r^4+2R^2A^2r^2+A^4R^2)\frac{dh_2}{dr}  \label{w2} \\
&&+3B^2r^3(A^2+2R^2)(A^2+r^2)\frac{dk_2}{dr}  \nonumber \\
&&-12B^2R^2(A^2+2r^2)^2(h_2+k_2)+A^2r^4(r^2-R^2)(A^2+2r^2)\left( \frac{%
d\omega }{dr}\right) ^2=0\ .  \nonumber
\end{eqnarray}
In the following, we use equations (\ref{ford1}) and (\ref{ford2}) to
eliminate $h_2$ and $m_2$, while we employ (\ref{tolmom}) to express the
second derivative of the rotation potential $\omega $. Comparing Eq. (\ref
{w2}) with the derivative of (\ref{w1}) we get another first order
differential equation. Eliminating the first derivative of $k_2$ by (\ref{w1}%
), we get an equation that can be solved algebraically for $k_2$.
Substituting $k_2$ back to (\ref{w1}) again, we obtain an equation
containing only $\omega $. Taking the $r$ derivative of this equation, we
readily find that it is not consistent with (\ref{tolmom}). Hence we have
proven

\textbf{Theorem 3:} A rotating perfect fluid spacetime that reduces to the
Tolman IV solution in the static limit and can be matched to an
asymptotically flat vacuum exterior cannot be Petrov type D.

\section{Acknowledgment}

This work has been supported by OTKA grant T022533. The authors would like
to thank M. Bradley and M. Marklund for fruitful discussions. G.\ F.\ would
like to acknowledge the support of the Japan Society for the Promotion of
Science and thank for the hospitality of the Physics Department of Waseda
University.

\end{document}